\documentstyle[aps,prl,epsf,epsfig]{revtex}

\newcommand{\nc}{\newcommand}
\nc{\beq}{\begin{equation}}
\nc{\eeq}{\end{equation}}
\nc{\beqa}{\begin{eqnarray}}
\nc{\eeqa}{\end{eqnarray}}
\nc{\lra}{\leftrightarrow}

\nc{\sss}{\scriptscriptstyle}
\nc{\PL}{P_{\sss L}}
\nc{\PR}{P_{\sss R}}
\nc{\lsim}{\mbox{\raisebox{-.6ex}{~$\stackrel{<}{\sim}$~}}}
\nc{\gsim}{\mbox{\raisebox{-.6ex}{~$\stackrel{>}{\sim}$~}}}

\begin{document}
\twocolumn[\hsize\textwidth\columnwidth\hsize\csname@twocolumnfalse%
\endcsname

\title{Higgs Mechanism and Bulk Gauge Boson Masses in the 
Randall-Sundrum Model}

\author{Stephan J.~Huber$^*$ \vspace{-.3cm}}
\author{Qaisar Shafi$^{\dagger}$ \vspace{.2cm}} 
\address{\it Bartol Research Institute, University of Delaware, Newark, DE 19716, USA}

\maketitle

\begin{abstract}
Assuming the breaking of gauge symmetries by 
the Higgs mechanism, we consider the associated bulk gauge
boson masses in the Randall-Sundrum background.   
With the Higgs field confined on the TeV-brane, the W and Z 
boson masses can naturally be an order of magnitude smaller than
their Kaluza-Klein excitation masses. Current electroweak precision 
data requires the lowest excited state to lie above about 30 TeV,
with fermions on the TeV-brane. This bound is reduced to about 
10 TeV if the fermions reside sufficiently close to the Planck-brane.
Thus, some tuning of parameters is needed.
We also discuss the bulk Higgs case, where the bounds
are an order of magnitude smaller.
\vspace{.2cm}
\end{abstract}

%
%
]

It has recently been realized that the large hierarchy between the 
Planck scale and the electroweak scale could be related to the 
presence of extra dimensions \cite{ADD}. An interesting realization
of this concept is the Randall-Sundrum model \cite{RS}. It relies
on the 5-dimensional non-factorizable geometry
\begin{equation} \label{met}
ds^2=e^{-2\sigma(y)}\eta_{\mu\nu}dx^{\mu}dx^{\nu}+dy^2,
\end{equation}
where $\sigma(y)=k|y|$.  
The 4-dimensional metric is $\eta_{\mu\nu}={\rm diag}(-1,1,1,1)$, 
$k$ is the AdS curvature, and $y$ denotes the fifth dimension.
This metric results from a suitable adjustment of the bulk cosmological
constant and the tensions of the two 3-branes which reside 
at the $S_1/Z_2$ orbifold fixed points $y=0$, $y=\pi R$.   
Because of the exponential (``warp'') factor, the effective mass scale
on the brane located at $y=\pi R$ is $M_P e^{-\pi kR}$. 
If $kR\sim11$ this scale will be ${\cal O}$(TeV), and the brane
is referred to as the `TeV-brane'.  Hence
the model can generate an exponential hierarchy of scales
from a small extra dimension.

In the setting of ref.~\cite{RS} only gravity propagates in
the 5d bulk, while the Standard Model (SM) fields are confined 
to the TeV-brane.
However, since a microscopic derivation is still missing,
it is interesting to study other possibilities. Bulk scalar
fields were first discussed in ref.~\cite{GW}. The consequences
of SM gauge bosons in the bulk were studied in ref.~\cite{DHR,P}.
In ref.~\cite{GN} the behavior of fermions in the bulk was investigated,
and in ref.~\cite{CHNOY} the complete SM was put in the bulk. 
Finally, bulk supersymmetry was considered in ref.~\cite{GP}. 

Bulk gauge fields are necessary if the SM fermions live in the bulk. 
By localizing the fermions at different positions in the fifth dimension  
it seems possible to address the questions of fermion mass hierarchy,
non-renormalizable operators and proton decay \cite{AS,DDG,GP}.
New possibilities for baryogenesis may open up
if the fermion separation is reduced by thermal correction
in the hot early universe \cite{MPST}.

Bulk vector bosons with bulk masses have been considered 
to some extent in refs.~\cite{P,CHNOY}. It was found that 
the `zero' mode acquires a mass comparable to the mass
of the `first' Kaluza-Klein (KK) excitation,  unless the bulk 
gauge boson mass is extremely fine-tuned \cite{CHNOY}. Since
gauge boson KK excitations should have masses in the TeV range \cite{DHR,P},
the W-boson mass could only be generated by reintroducing the
original hierarchy problem. 
This suggests \cite{CHNOY} that the Higgs should be confined to the
TeV-brane, i.e. the gauge boson mass arises from the boundary.  

In this letter we will investigate this scenario in more detail. 
We will study the properties of bulk gauge bosons
which are related to broken gauge symmetries, i.e. bulk 
W and Z bosons. We will show
that in the case of a TeV-brane Higgs field, the W and Z boson masses 
are naturally an order of magnitude smaller
than the mass of their first KK excitations. We will demonstrate that 
the W and Z boson mass ratio and  $\sin^2\theta_W$
can be successfully reproduced by a moderate tuning of the brane
mass parameter. We also discuss constraints from 
universality of the coupling of the gauge bosons to fermions. 
In the phenomenologically viable parameter range we recover 
the 4d relationship between gauge and Higgs boson masses.
Contraints arising in the bulk Higgs case are also briefly
discussed.   
%

Let us consider the following equation of motion for a U(1) gauge 
boson $A_N$ of 5-dimensional mass $M$
\begin{equation} \label{EL}
\frac{1}{\sqrt{-g}}\partial_M(\sqrt{-g}g^{MN}g^{RS}F_{NS})-M^2g^{RS}A_S=0,
\end{equation}
where $g_{MN}$ denotes the 5-dimensional metric.
In general, $M$  arises from some Higgs
mechanism and consists of bulk and boundary 
contributions
\begin{equation}\label{mass}
M^2(y)=b^2k^2+a^2k\delta(y-\pi R) + \tilde a^2k\delta(y),
\end{equation}
depending on whether the Higgs fields live in the bulk 
and/or on the branes. 
The gauge boson masses can be expressed in terms of the 
parameters of the Higgs potential. For the TeV-brane Higgs, for
instance,  we have 
\begin{equation}\label{a}
a^2=\frac{g_5^2\mu^2}{2\lambda k}=\frac{g_5^2}{2k}v^2,
\end{equation}  
where $\mu$ denotes the Higgs mass parameter and $\lambda$ the quartic 
coupling, both understood as 4d quantities, and $g_5$ is the 5d gauge 
coupling. $v=\mu/\sqrt{\lambda}$ is the vev of the Higgs field.

Using the metric 
(\ref{met}) and decomposing the 5d field as 
\begin{equation}\label{KK}
A_{\mu}(x^{\mu},y)=\frac{1}{\sqrt{2\pi R}}\sum_nA_{\mu}^{(n)}(x^{\mu})f_n(y),
\end{equation}
one obtains \cite{DHR,P} 
\begin{equation}\label{eqm}
(-\partial_y^2+2\sigma'\partial_y+M^2)f_n=e^{2\sigma}m_n^2f_n,
\end{equation}
where $m_n$ are the masses of the Kaluza-Klein 
excitations $A_{\mu}^{(n)}$,
and $\sigma'=\partial_y\sigma$. (We work in the gauge $A_5=0$ 
and $\partial_{\mu}A^{\mu}=0$.)

Requiring the gauge boson wave function to be even under the $Z_2$
orbifold transformation, $f_n(-y)=f_n(y)$, one finds \cite{P}
\begin{equation}\label{wave}
f_n=-\frac{e^{\sigma}}{N_n}\left[J_{\alpha}(\frac{m_n}{k}e^{\sigma})+
  \beta_{\alpha}(m_n)Y_{\alpha}(\frac{m_n}{k}e^{\sigma})\right],
\end{equation}
where the order of the Bessel functions is $\alpha=\sqrt{1+b^2}$. 
The coefficients $\beta_{\alpha}$ obey
\begin{eqnarray} \label{quant1}
 \beta_{\alpha}(x_n,\tilde a^2)&=&-\frac{(-\frac{\tilde a^2}{2}+1-\alpha)
                            J_{\alpha}(x_n)
                            +x_nJ_{\alpha-1}(x_n)}
                           {(-\frac{\tilde a^2}{2}+1-\alpha)Y_{\alpha}
                            (x_n)
                            +x_nY_{\alpha-1}(x_n)}, 
\\[2mm]  \label{quant2}
 \beta_{\alpha}(x_n,\tilde a^2)&=&\beta_{\alpha}(\Omega x_n,-a^2),
\end{eqnarray}
where we defined the warp factor $\Omega=e^{\pi kR}$, 
and $x_n=\frac{m_n}{k}$.
Note that for non-vanishing boundary mass terms the derivative
of $f_n$ becomes discontinuous on the boundaries. The normalization
constants $N_n$ are defined such that $(1/\pi R)\int_0^{\pi R}dyf_n^2=1$.
%

An analogous discussion also holds in the non-abelian case.

Eqs.~\ref{quant1}, \ref{quant2} encode the masses of the different KK states.
In the limit $m_n \ll k$ and $m_n \Omega \gg k$, one finds \cite{GP}
\begin{equation}
m_n\approx(n+\frac{\alpha}{2}-\frac{3}{4})\pi k \Omega^{-1}.
\end{equation} 
In this regime the masses of the excited KK states are independent of the 
boundary mass terms. The bulk mass term enters via $\alpha$, but
its contribution is also suppressed by the warp factor. 
This is because the excited states are localized at the TeV-brane 
as a result of the exponential in their wave functions.
If the SM fermions live on the TeV-brane, it was found that the 
masses of the gauge boson KK states should 
be in the multi TeV range in order to be in agreement with
the electroweak precision data \cite{DHR}. In the case of
bulk fermions the corresponding constraints
becoming weaker \cite{CHNOY,GP}, reducing to $m_1>0.5$ TeV
for fermions on the Planck-brane \cite{P}.  
 
Let us now consider $m_0$, the mass of the lowest lying state.
In the case with neither bulk nor boundary mass term, one 
finds $m_0=0$, and the corresponding (zero mode) wave
function is not localized in the extra dimension, $f_0(y) \equiv 1$.
If a bulk or boundary mass term is added, the `zero' mode
picks up a mass, and its wave function displays a y-dependence.

In the case of a bulk mass term $b\sim 1$, one 
finds $m_0\sim \pi k \Omega^{-1}$, 
i.e.  approximately of the same value as the first excited 
state in the massless case \cite{CHNOY}. 
Although the bulk mass term is of order $M_p$,
the gauge boson mass does not become Planck-sized, because
$f_0$ is localized at the TeV-brane, where the effective mass 
scale is small. At first sight this seems encouraging, but it was also
shown in ref.~\cite{CHNOY} that extreme fine tuning $b\sim\Omega^{-1}$
is necessary in order to bring down the W-boson mass,
i.e. $m_0$, from its natural TeV size range to the experimental value.       
One therefore would have to start with a weak scale bulk mass term, 
which is nothing but the original fine tuning problem.
These results follow from expanding eqs.~\ref{quant1}, \ref{quant2} in the
regime $x_n\ll 1$ and $x_n\Omega\ll 1$. Along the same lines we find
that a gauge boson mass term at the Planck-brane has the same
implications, 
\begin{equation}
x_0^2\approx \frac{\tilde a^2}{2\ln \Omega}=\frac{\tilde a^2}{2\pi kR}.
\end{equation}
Since their is no warp factor suppression, only for 
$\tilde a\sim \Omega^{-1}$ is a W-mass below the Kaluza-Klein
scale possible.

If the Higgs is on the TeV-brane however, we arrive at a different 
conclusion. Expanding eqs.~\ref{quant1}, \ref{quant2} for $x_0\ll 1$ 
and $\Omega x_0 \ll 1$ we find 
\begin{equation} \label{1}
\Omega^2x_0^2\approx \frac{a^2}{2\ln\Omega}\mbox{ }
   \frac{1}{1-\frac{a^2}{4\ln\Omega}
  (\gamma+\ln\frac{x_0}{2})},
\end{equation}
where $\gamma\approx 0.5772$, which reduces to 
\begin{eqnarray} \label{2}
\Omega^2x_0^2&\approx&\frac{a^2}{2\ln\Omega}=\frac{a^2}{2\pi kR}, \quad a\ll1
\\ \label{3}
\Omega^2x_0^2&\approx&\frac{2}{\ln\Omega}=\frac{2}{\pi kR}, \quad a\gg1.
\end{eqnarray}
Similar to the case of a bulk or Planck-brane mass term, we 
find a linear relationship between $a$ 
and $x_0$ for small values of $a$. But in contrast to the former, this
behavior remains valid up to $a\lsim 1$, because of the appearance of the
warp factor. For $a\gsim 1$, $x_0$ saturates at a value typically an order
of magnitude smaller than $x_1$, which corresponds to the mass
of the first excited state. This demonstrates that a Higgs boson at
the TeV-brane can, in principle, explain weak gauge boson masses
of order 100 GeV, while keeping the KK states 
in the TeV range \cite{f1}.
The saturation results from the drop of the wave function near the
TeV-brane for large $a$  which diminishes the 
overlap with the brane mass term.

\begin{figure}[b] 
\begin{picture}(100,115)
\put(95,-55){\epsfxsize4cm \epsfbox{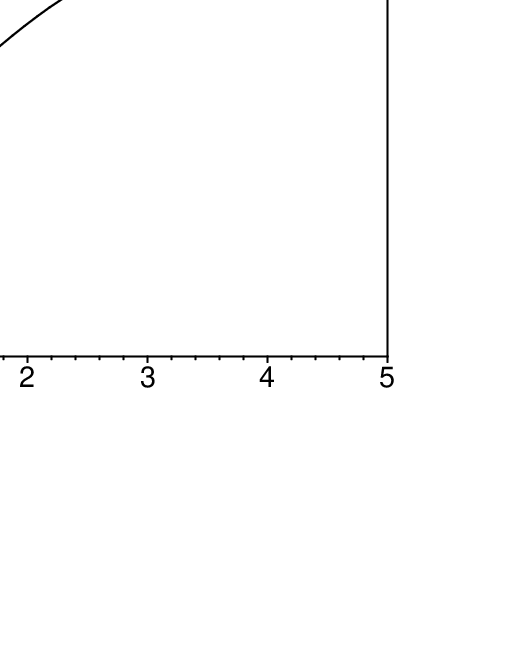}}
\put(112,0){{\footnotesize $a$}}
\put(12,69){{\footnotesize $\Omega x_0$}}
\end{picture} 
\caption{$\Omega x_0$ versus $a$ for a warp factor 
$\Omega=10^{14}$.}
\label{f_1}
\end{figure}

\begin{figure}[b] 
\begin{picture}(100,110)
\put(95,-46){\epsfxsize4cm \epsfbox{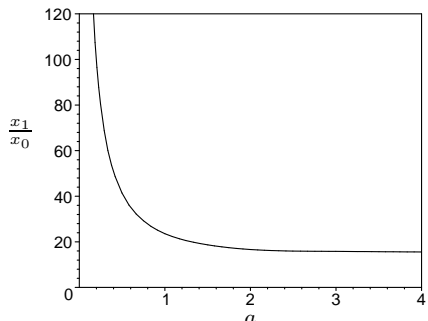}}
\put(113,3){{\footnotesize $a$}}
\put(23,74){{\footnotesize $\frac{x_1}{x_0}$}}
\end{picture} 
\caption{Mass ratio of the lowest lying and  the first excited
KK state versus $a$ for $\Omega=10^{14}$.}
\label{f_2}
\end{figure}

In fig.~\ref{f_1} we show $\Omega x_0$ as a function of $a$ 
for $\Omega=10^{14}$, i.e.~$kR=10.26$. For $a\gg1$ we
obtain $\Omega x_0 \sim 0.24$. 
(In the evaluation we numerically solved eqs.~\ref{quant1}, \ref{quant2},
but eq.~\ref{1} would also reproduce the results at a percent
accuracy level.)
The mass of the first excited KK state
depends very weakly on $a$. In our example we find it
rises from $\Omega x_1(a=0)=2.45$ \cite{DHR,P} to $\Omega x_1(\infty)=3.88$.
In fig.~\ref{f_2} we display the resulting ratio between the
mass of the ground state and the first excited state. For small 
enough $a$ we find $x_1/x_0\equiv m_1/m_0\propto1/a$, while for large $a$
the ratio approaches $x_1/x\sim 15.4$. For different 
values of $\Omega$ the results hardly change since the 
warp factor only enters logarithmically in (\ref{2}, \ref{3}).

Since the ground state
mass scale, i.e. the W and Z boson masses, is experimentally known
to be $\sim 100$ GeV, we conclude that in the brane Higgs
scenario $m_1\gsim1.5$ TeV is necessary. This bound
does not rely on the electroweak precision data and 
is independent of the position of the fermions in the
fifth dimension. It could only be weakened if the warp
factor in (\ref{3}) is substantially reduced, which
would reintroduce the hierarchy problem.


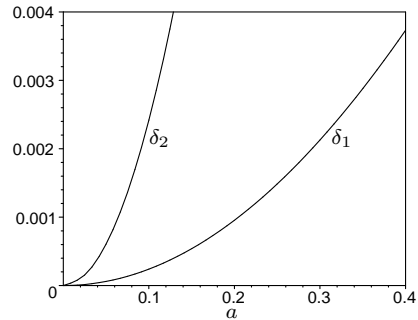
\begin{figure}[t] 
\begin{picture}(100,120)
\put(95,-50){\epsfxsize4cm \epsfbox{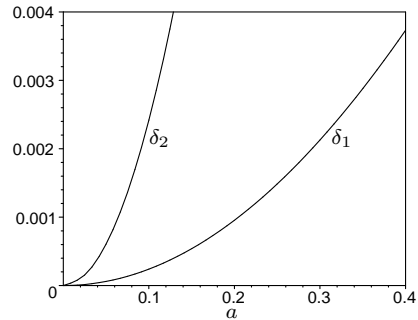}}
\put(112,4){{\footnotesize $a$}}
\put(152,70){{\footnotesize $\delta_1$}}
\put(83,70){{\footnotesize $\delta_2$}}
\end{picture} 
\caption{Plots of $\delta_1$ (see eq.~\ref{d1})
and $\delta_2$ (\ref{d2}) versus $a$, with $\Omega=10^{14}$.}
\label{f_3}
\end{figure}

%
We next discuss constraints on KK excitations arising 
from the electroweak precision data.
From fig.~\ref{f_1} we deduce that the relationship between
the boundary mass term $a$ and the ground state mass
becomes highly non-linear in the regime $a\gsim1$. As a
result the very successful SM prediction that the gauge
boson masses are proportional to their couplings to the 
Higgs could be spoiled. We measure the deviation from the
linear behavior of $x(a)$ by
\begin{equation} \label{d1}
\delta_1\equiv\frac{x_0(ar)}{x_0(a)}-r,
\end{equation}  
and take $r=M_W/M_Z=0.88$. For $\Omega=10^{14}$ the results
are shown in fig.~\ref{f_3}. They are well approximated 
by $\delta_1\sim 0.025a^2$, i.e. the non-linearity increases
quadratically with $a$.  Since $r$ is measured to an
accuracy of about $10^{-3}$ 
and no deviations from the SM prediction have been found \cite{pdg},
we require $\delta_1\lsim10^{-3}$. This leads to the 
modest constraint $a\lsim 0.2$. From fig.~\ref{f_2} we deduce that
$x_1/x_0\gsim 100$. As a result the mass of the first KK excitation
has to obey $m_1\gsim 10$ TeV. The constraint on $m_1$ is proportional 
to $1/\sqrt{\delta_1}$. We stress again that it
does not depend on where the fermions live. The warp factor 
only enters logarithmically. 

\begin{figure}[t] 
\begin{picture}(100,105)
\put(95,-52){\epsfxsize4cm \epsfbox{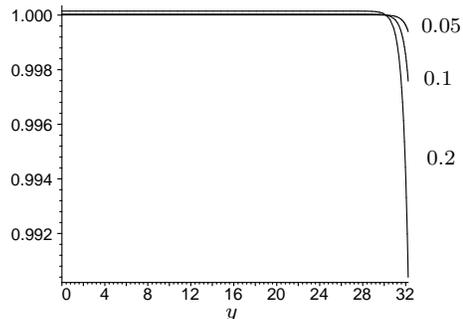}}
\put(112,2){{\footnotesize $y$}}
\put(188,60){{\footnotesize $0.2$}}
\put(187,90){{\footnotesize $0.1$}}
\put(186,110){{\footnotesize $0.05$}}
\end{picture} 
\caption{Ground state wave function $f_0(y)$ 
(eq.~\ref{wave}) for $a=0.2$, 0.1, 0.05, where
$\Omega=10^{14}$.}
\label{f_6}
\end{figure}
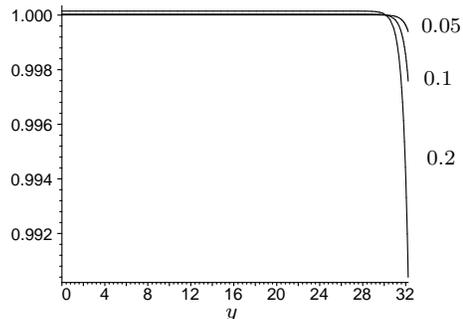 

Once the `zero' mode acquires a mass its wave function (\ref{wave}) 
depends on the $y$ coordinate. In contrast to excited states
the wave function
tries to avoid the TeV-brane where its mass arises, as 
shown in fig.~\ref{f_6}. As a result the successful SM predictions 
of the gauge couplings to fermions of the W and Z bosons 
can be affected. The resulting constraints depend on
the position of the fermions in the fifth dimension.

For example, the coupling of the W boson to a fermion on 
the TeV-brane is given by $g_0f_0(\pi R)$, where $g_0$
denotes the coupling if the boson were massless.
Since $f_0(\pi R)<1$ in the brane Higgs scenario, the resulting
gauge coupling is somewhat reduced. In fig.~\ref{f_3} we present
the resulting deviation from the SM prediction,
\begin{equation} \label{d2}
\delta_2\equiv1-f_0(\pi R),
\end{equation}     
as a function of the brane mass parameter $a$ ($\Omega= 10^{14}$).   
For $\delta_2\lsim10^{-3}$, we find $a\lsim 0.06$, a constraint
more stringent than the one from the mass ratio $r=M_W/M_Z$. From 
fig.~\ref{f_2} we learn that $x_1/x_0\gsim 310$, i.e. $m_1\gsim 30$ TeV,
a bound which is proportional to $1/\sqrt{\delta_2}$.
With this restriction the effects of the KK states are automatically 
in agreement with the electroweak 
precision data, which only requires $m_1\gsim 23$ TeV \cite{DHR}.

If the massive gauge boson is coupled to fermions on the Planck-brane
the effective gauge couplings hardly deviate from the SM prediction,
since $f_0(0)\sim1$ (see fig.~\ref{f_6}). 
The resulting constraint, $m_1 \gsim 4$ TeV, is weaker than that
arising from eq.~\ref{d1}. 

\begin{figure}[t] 
\begin{picture}(100,120)
\put(95,-50){\epsfxsize4cm \epsfbox{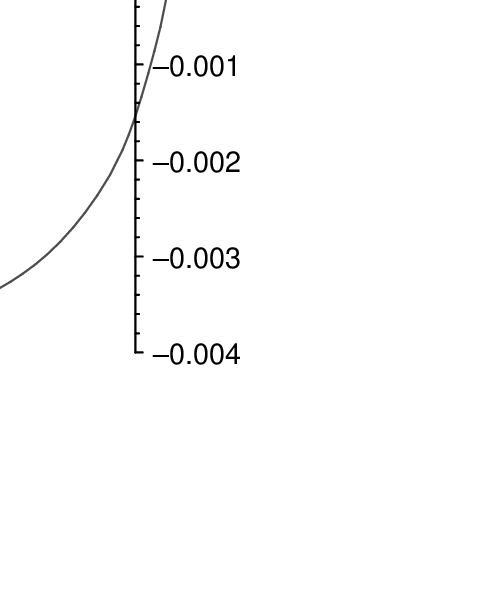}}
\put(190,95){{\footnotesize $c$ $\rightarrow$}} 
\put(95,115){{\footnotesize $\frac{g-g_0}{g_0}\uparrow$}}
\end{picture} 
\caption{Deviation of $g/g_0$ from unity versus the
fermion mass parameter $c$, for $a=0.14$ and $\Omega=10^{14}$.}
\label{f_5}
\end{figure}

Bulk fermions interpolate between the TeV-  
and the Planck-brane scenarios. As discussed in refs.~\cite{GN,GP},
depending on the bulk mass term $m_{\psi}=c\sigma'$ for the fermion,
the zero mode of the fermion is localized at the TeV-brane ($c<1/2$) 
or at the Planck-brane ($c>1/2$). For $c=1/2$ the fermionic
zero mode is delocalized in the fifth dimension. Since the W-boson
wave function has a nontrivial $y$-dependence, it then couples 
non-universally
to fermions localized at different positions in the fifth dimension. This
is completely analogous to the $c$-dependent
coupling of the excited gauge boson states discussed in 
ref.~\cite{GP}. 
We have repeated the analysis for the ground state of the massive
gauge boson. In fig.~\ref{f_5} we display $g/g_0-1$ as function of $c$ 
for $a=0.14$ and $\Omega=10^{14}$, where $g_0$ would be the coupling of a
massless gauge boson. The shape of the gauge coupling of the massive
ground state is similar to those of the excites KK states \cite{GP}. 
However, the amplitude of the variation is much smaller. 
In the limit $c\rightarrow -\infty$,
$g$ approaches the result of the TeV-brane fermions (\ref{d2}).
In the regime $c\gsim 1/2$ the deviation of the SM prediction for $g$ becomes
small. In this case $a$ is only constrained by the W,Z boson mass ratio
(\ref{d1}).
 
If the SM fermions reside on the TeV-brane, non-renormalizable
operators are typically suppressed only by a few TeV instead
of the large 4d Planck mass \cite{DDG}. This may induce rapid 
proton decay, large flavor changing neutral currents, and large 
neutrino masses. Global symmetries like baryon and and number
may be imposed to forbid the corresponding operators. If the SM
fermions are bulk fields, the non-renormalizable interactions
can be suppressed to some extent by localizing the first and
second generation fermions near the Planck-brane \cite{GP}. 
We will address this topic in a future publication \cite{HS}.

%
Taking into account the warp factor, the Higgs mass on the 
TeV-brane is given by \cite{RS}
\begin{equation}
M_H=\mu\Omega^{-1}=\sqrt{\lambda}v\Omega^{-1}.
\end{equation}
In 4d the gauge and Higgs boson masses are related by 
$M_W^2=\frac{g^2}{2\lambda}M_H^2$. In the brane Higgs
scenario this relationship is certainly violated in the
regime $a\gsim 1$ due to the non-linearity in $m_0(a)$
(see fig.~\ref{f_1}). However, in the phenomenologically
viable parameter range $a\lsim a_{\rm max}\approx 0.14$, where 
$m_0^2\approx (g_5^2/4\pi R)v^2$
(\ref{2}, \ref{a}), and $g\approx g_5/\sqrt{2\pi R}$, the 4d 
relation is recovered, up to small correction of 
order $10^{-3}$. Using (\ref{a}), the parameters of
the Higgs potential have to obey
\begin{equation} 
\frac{\mu^2}{k^2}<\frac{a_{\rm max}^2}{2\pi kR}\frac{2\lambda}{g^2}.
\end{equation}
Assuming $\lambda\sim1$ we find that a moderate tuning $\mu\lsim0.04k$
is required to reproduce the measured W and Z boson masses in the 
brane Higgs scenario. 

%
Finally, let us briefly summarize our results for the bulk Higgs case, 
which may be especially interesting if SM fermions reside on the 
Planck-brane in 
order to eliminate unwanted higher dimensional operators.
A TeV-brane Higgs cannot provide masses for these fermions.
To solve the hierarchy problem, one has to rely on some 
additional mechanism, for instance supersymmetry.
The W,Z boson mass ratio (\ref{d1}) leads to 
$m_1\gsim 250$ GeV, which is rather weak compared to the  
brane Higgs case ($m_1\gsim 30$ TeV). Stronger restrictions arise from the 
modification of the gauge couplings (\ref{d2}). For Planck-brane 
fermions we find $m_1 \gsim 600$ GeV, while for TeV-brane
fermions this increases to $m_1 \gsim 3.5$ TeV, bounds which are
again weaker than for the TeV-brane Higgs case. The wave function
of the `zero' mode increases near the TeV-brane. This results
also apply in the Planck-brane Higgs scenario.

\vspace{0.1in}
We would like to thank G.~Dvali and G.~Gabadadze for very useful
discussions.
S.~H.~is supported in part by the Alexander von Humboldt foundation.
This work was also supported by DOE under contract DE-FG02-91ER40626. 
\vspace{0.3in}
\\
$^*$email address: shuber@bartol.udel.edu
\\
$^{\dagger}$email address: shafi@bartol.udel.edu


\begin{references}
\bibitem{f1} If we consider a scalar field instead of
             the vector field, eq.~\ref{1} becomes 
          $\Omega^2x_0^2\approx \Omega^{-2}a^2/(1+\frac{a^2}{8})$, i.e. 
         the generated scalar mass is ${\cal O}(\Omega^{-1}$TeV), which
         is of sub-eV size.
\end{references}
\end{document}